\documentstyle[aps,preprint]{revtex}

\begin{document}

\preprint{cond-mat/9309030}

\draft

\title{Exact Dynamical Correlations Of The $1/r^{2}$ Model}
\author{E.~R.~Mucciolo$^{(a)}$, B.~S.~Shastry$^{(b)}$,
B.~D.~Simons$^{(a)}$, and B.~L.~Altshuler$^{(a)}$,}
\address{$^{(a)}$Department of Physics, Massachusetts Institute of
Technology, \\ Cambridge, Massachusetts 02139 \\ $^{(b)}$AT$\&$T Bell
Laboratories, Murray Hill, New Jersey 07974}

\maketitle

\begin{abstract}
We present exact results for the dynamical structure function,
i.e.~the density-density correlations for the $1/r^2$ system of
interacting particles at three special values of the coupling
constant. The results are interpreted in terms of exact excitations of
the model which are available from Bethe's Ansatz, thereby throwing
light on the quasi-particle content of the elementary excitations. We
also obtain the first moment of the discrete version of the model,
which provides a non-trivial check on its structure function. We show
that the property of spectrum saturation is a common feature of both
versions.
\end{abstract}


\section{Introduction}
\label{sec:intro}

The $1/r^2$ system of interacting particles, introduced by Calogero
\cite{Calogero} and Sutherland \cite{Sutherland}, and indeed even
earlier by Dyson \cite{Dyson}, within a relaxational framework in the
course of his discussion of the Brownian motion of Random Matrices,
has continued to be of great interest. In the recent past, it has also
generated great interest in the context of a discrete version, i.e. a
spin $1/2$ model introduced independently by Haldane and Shastry
\cite{Haldane,Shastry88}, and in terms of its algebraic content
\cite{Sh-Su-90,Sutherland92,Shastry92,Sh-Su-93,Su-Sh-93}. Its
interest derives from the combination of beautiful
mathematical structure and rich physical phenomena of Quantum
Fluctuations in a low dimensional system (quasi-LRO, non Fermi Liquid
behavior etc.), as well as surprising tractability.

In a recent development in an apparently completely different physical
system, namely that of electrons in a random medium, Simons {\it et
al.} \cite{Ben_Lee_Boris_PRL,Ben_Lee_Boris_PRB} have succeeded in
computing a certain correlation function depending on two variables,
say space and time, and conjectured that this represents the
density-density correlation function of the above $1/r^2 $ model at
three appropriate values of the coupling constant $\beta=$ 1, 2, and
4, corresponding to orthogonal, unitary, and symplectic ensembles,
respectively. The results are obtained due to a suggested equivalence
of the problem to the evolution of energy eigenvalues of a disordered
metallic grain subject to an arbitrary perturbation \cite{Ben_Boris}
to that of the $1/r^2 $ many body problem. The mapping, performed on
the level of the two-point, time-dependent, density-density
correlation functions, leads to an explicit exact result for the
density-density correlation of the $1/r^2$ model for the above values
of the coupling constant. The result, astonishingly enough, is valid
at {\it all length and energy scales}, not just in asymptotic regions.
This conjecture has been explicitly confirmed in a recent work by
Narayan and Shastry \cite{Narayan}, where they established the
correspondence between the evolution of the distribution of
eigenvalues of a random matrix subject to a random Gaussian
perturbation, and a Fokker-Planck equation which is equivalent to the
$1/r^2$ model. At the same time, Simons {\it et al.}
\cite{Ben_Lee_Boris_NPB} have established a direct connection between
the $1/r^{2}$ model and correlations in the spectra of random matrices
through a continuous matrix model.

The excitation spectrum of the $1/r^2$ model is available in great
detail from the (Asymptotic) Bethe's Ansatz (ABA) invented by
Sutherland \cite{Sutherland_course}. The picture that arises is that
of an underlying gas consisting of quasi-particles obeying Fermi
statistics, and interacting weakly with each other through a
Hartree-Fock interaction leading to a back flow. This picture indeed
gives {\it all} the excited states of the model. The remaining problem
then, is that of an appropriate decomposition of the ``bare''
particles into ``quasi-particles''. The explicit knowledge of the
correlation functions provides us with an opportunity to describe the
intermediate states in $S(q,\omega ) $ phenomenologically as
combinations of the ``quasi-particles'', whose energies are available
from the ABA. This is analogous to quark spectroscopy in the theory of
elementary particles, the ABA quasi-particles and quasi-holes are our
quarks in the present scheme. We also provide a nontrivial check on
the conjecture by comparing the explicitly known functions of the
momentum $q$ of the $1/r^2$ model, with those obtained by integrating
the explicit correlation functions.

The plan of the paper is as follows. In Section \ref{sec:mapping}, we
define the $1/r^2$ model, and summarize the known information about
its moments. The dynamical correlations found in
Refs.~\cite{Ben_Lee_Boris_PRL,Ben_Lee_Boris_PRB} are summarized and
their Fourier transforms given. The small $q$ Hydrodynamic limit of
these correlators is calculated, and the saturation of the spectrum by
a sound like linear mode (i.e. $\omega = s q$) is demonstrated.
Section \ref{sec:ABA} contains a summary of the results of the
Asymptotic Bethe's Ansatz for the $1/r^2$ model, where we write down
the dispersion of the effective quasi-particles and quasi-holes. In
Section \ref{sec:content}, we rework the expressions for the structure
function for the three ensembles into forms wherein the energy
conserving delta functions are shown to have a natural interpretation
in terms of multi quasi particle-hole pairs. In Section
\ref{sec:discrete}, we discuss the discrete $1/r^2$ model, and display
its first three relevant moments explicitly. The similarity to the
continuum $1/r^2$ model in terms of the exhaustion of the structure
function by ``spinons'' at low $q$ is pointed out, and this is
highlighted to be a unique feature characterizing this family of
models. In Section \ref{sec:conclusions}, we summarize our results.

\section{Universal Correlation Functions Applied To The $1/r^{2}$ Model}
\label{sec:mapping}

The Sutherland-Calogero-Moser system with periodic boundary conditions
has a Hamiltonian describing spinless fermions confined to a ring and
interacting through a $1/r^{2}$ pairwise potential \cite{Sutherland}:
\begin{equation}
H = -\sum_{i=1}^{N}\frac{\partial^{2}}{\partial r_{i}^{2}} \ + \
\beta(\beta/2-1)\sum_{i>j}
\frac{(\pi/L)^{2}}{\sin^{2}[\pi(r_{i}-r_{j})/L]} \ .
\label{eq:hamiltonian}
\end{equation}
For simplicity we have chosen the mass to be 1/2. The ring has length
$L$ and the number of particles is $N$. The statistics of the
particles can be chosen arbitrarily since the particles cannot get
past each other owing to the singular nature of the interaction at the
origin; we declare them to be fermions for convenience (they are
indeed so at $\beta=2$), and one might imagine the system to be that
of fermions with either repulsive ($\beta > 2$) or attractive ($\beta
< 2$) interactions. Unless explicitly specified, we will assume that
the system is in the thermodynamic limit ($L\rightarrow\infty$ and
$N\rightarrow\infty$) and has a finite $O(1)$ density ($d=N/L$).

It was first argued in Ref. \cite{Ben_Lee_Boris_PRL} that the
time-dependent correlation functions of this one-dimensional
Hamiltonian are equivalent to certain universal correlation functions
\cite{Ben_Boris} of the energy spectra of weakly disordered metallic
grains when $\beta=1$, 2, or 4. For spectra subject to some arbitrary
perturbation, $X$, exact analytical expressions were derived for the
two-point density correlation function,
\begin{equation}
k(E,X) = \langle\rho(\bar{E}-E,\bar{X}+X)\rho(\bar{E},\bar{X})\rangle
- \langle\rho(\bar{E},\bar{X})\rangle^{2} \ ,
\label{eq:correl1}
\end{equation}
where $\rho(\bar{E},\bar{X}) =
\sum_{i=1}^{N}\delta(\bar{E}-E_{i}(\bar{X}))$ is the density of states
of the system and $\langle \cdots \rangle$ denotes a statistical
average which can be performed over a range of energy or over $X$. It
was shown that after the following rescaling in which the parameters
become dimensionless, $\epsilon_i=E_i/\Delta$ and $x=X \sqrt{\langle
(\partial\epsilon_i(X)/\partial X)^2\rangle}$, Eq. (\ref{eq:correl1})
becomes universal, depending only on the symmetry of the Dyson ensemble.
In fact, the universality is not specific to disordered metals but
applies equally to all non-integrable or quantum chaotic systems
\cite{Ben_Boris}.

Remarkably, by performing the change of variables
\begin{equation}
x^{2} \equiv -2it \ , \epsilon \equiv r \ ,
\label{eq:mapping}
\end{equation}
where $t$ is the time coordinate and $r$ is the spatial coordinate,
Eq. (\ref{eq:correl1}) becomes equivalent to the two-point particle
density correlator of the ground state of the Sutherland model. The
coordinate $r$ will be given in units of the mean interparticle
distance $1/d$ (although $d \equiv \langle\rho\rangle=1$, we will
continue to display the ``d'' dependence in order to retain
generality). The resulting correlation function, after the change of
variables, is
\begin{equation}
k(r,t) = \langle\rho(\bar{r}-r,\bar{t}+t)\rho(\bar{r},\bar{t} \
)\rangle - d^2 \ ,
\label{eq:correl2}
\end{equation}
with $\rho(r,t)=\sum_{i=1}^{N}\delta(r-r_{i}(t))$. The expectation
value $\langle\cdots\rangle$ is to be taken on the ground state of $H$.

Once we have an expression for $k(r,t)$, we can calculate the
dynamical structure factor $S(q,\omega)$ \cite{Pines}. The explicit
connection between $k(r,t)$ and $S(q,\omega)$ is made by taking
the space and time Fourier transforms of $k(r,t)$,
\begin{eqnarray}
S(q,\omega) & = & \frac{1}{2\pi d} \int dr \int dt \ k(r,t) \
e^{-i(qr-\omega t)} \nonumber \ .
\end{eqnarray}
$S(q,\omega)$  has a representation in terms of
the excited states of the system:
\begin{equation}
S(q,\omega) \equiv \frac{1}{N} \sum_{\nu\neq 0}
|\langle\nu|\rho_{q}|0\rangle|^{2} \delta(\omega-E_{\nu}+E_{0}) \ ,
\label{eq:sqw}
\end{equation}
where $H|\nu\rangle=E_{\nu}|\nu\rangle$, and
\begin{equation}
\rho_{q}=\int dr \ \rho(r) \ e^{-iqr} = \sum_{i=1}^{N} e^{-iqr_{i}} \ .
\label{eq:rho}
\end{equation}

In the following, we will present exact analytical expressions and
discuss some of the properties of $S(q,\omega)$ for the three special
values of $\beta$.

\subsection{The Moments of $S(q,\omega)$}

We begin by stating some important sum rules \cite{Pines} concerning
the function $S(q,\omega)$. Defining the moments of this function as
\begin{equation}
I_{n}(q) \equiv \int_{0}^{\infty} d\omega \ \omega^{n} \ S(q,\omega) \
,
\label{eq:moments}
\end{equation}
it follows from the velocity independence of the interaction that
\begin{equation}
I_{1}(q) = q^{2} \ .
\label{eq:f-sumrule}
\end{equation}
This is the statement of particle conservation and is the familiar
$f$-sum rule.  Another sum rule follows from the compressibility
relation\cite{Pines}:
\begin{equation}
\lim_{q\rightarrow 0} I_{-1}(q) = \frac{1}{s^{2}} \ ,
\label{eq:compresum1}
\end{equation}
where $s^{2}=2(\partial P/\partial d)$ is the square of the sound
velocity and $P$ is the pressure. However, since $P=-(\partial
E/\partial L)$ and for the Sutherland model the ground state energy is
known to be $E_{0}=(\pi^{2}\beta^{2}/12)(N^{3}/L^{2})$
\cite{Sutherland}, it follows that Eq.~(\ref{eq:compresum1}) can also
be written as
\begin{equation}
\lim_{q\rightarrow 0} I_{-1}(q) = \frac{1}{\pi^{2}d^{2}\beta^{2}} \ .
\label{eq:compresum2}
\end{equation}

Finally, we note that the zeroth moment, $I_{0}(q)$, also called the
static form factor (usually denoted by $S(q)$), has been determined
before for all three values of $\beta$ from Random Matrix Theory (RMT)
\cite{Mehta}. The connection between the $1/r^2$ model in the static
limit and the distribution of eigenvalues of a random matrix was
established by Sutherland in his early papers \cite{Sutherland}.
Therefore, from the results of Ref.~\cite{Sutherland} we anticipate
the following expressions:
\begin{equation}
I_{0}^{u}(q) = \frac{1}{2k_{F}}[|q| + (2k_{F} -
|q|)\theta(|q|-2k_{F})] \ \ \ \ \ \ (\beta=2) \ ,
\label{eq:I0u}
\end{equation}
\begin{eqnarray}
I_{0}^{o}(q) & = & \cases{ \frac{|q|}{k_{F}} \left[ 1 -\frac{1}{2}\ln
\left( 1+\frac{|q|}{k_{F}}\right) \right] \ , & $|q|< 2k_{F}$ \cr 2
-\frac{|q|}{2k_{F}} \ln \left| \frac{|q|+k_{F}}{|q|-k_{F}} \right| \ ,
& $|q|> 2k_{F}$ \ \ \ \ ($\beta=1$) \ , \cr}
\label{eq:I0o}
\end{eqnarray}
and
\begin{eqnarray}
I_{0}^{s}(q) & = & \cases{ \frac{|q|}{4k_{F}}\left[ 1 -\frac{1}{2}\ln
\left|1-\frac{|q|}{2k_{F}}\right| \right] \ , & $|q|<4k_{F} $ \cr 1 \
, & $|q|>4k_{F} $ \ \ \ \ ($\beta=4$) \ . \cr}
\label{eq:I0s}
\end{eqnarray}
For small $q$ we find the limiting behavior
\begin{equation}
I_{0}(q) \stackrel{q\rightarrow 0}{\longrightarrow} \frac{q}{\beta
k_{F}}.
\label{eq:limithydro}
\end{equation}

In terms of the three moments given here, we can calculate two
characteristic frequencies that give us an idea of the dispersion
relations of the excited modes, namely the Feynman
spectrum\cite{Feynman},
\begin{equation}
\omega_{\cal F}(q) \equiv \frac{I_{1}(q)}{I_{0}(q)} =
\frac{q^{2}}{S(q)} \ ,
\label{eq:omegaF}
\end{equation}
which has the small $q$ behavior
\begin{equation}
\omega_{\cal F}(q) \stackrel{q\rightarrow 0}{\longrightarrow}
\beta q k_F \ ,
\label{eq:omegaF_limit}
\end{equation}
and another one we call the ``hydrodynamical spectrum'',
\begin{equation}
\omega_{\cal H}(q) \equiv \sqrt{\frac{I_{1}(q)}{I_{-1}(q)}}
\stackrel{q\rightarrow 0}{\longrightarrow} \ \beta q k_{F} \ .
\label{eq:omegaH}
\end{equation}
In Fig.~1 we have plotted these dispersion relations for the three
values of $\beta$ for which we know the moments exactly. For the
repulsive and noninteracting cases there is a logarithmic dip at
$q=2k_{F}$, while for the attractive case the dispersion grows
monotonically. The appearance of a dip can be interpreted as a
tendency towards ``crystallization'' (i.e. the particles tend to
arrange themselves in a lattice with spacing $1/d$) as the strength of
the repulsive interaction increases.

\subsection{Static Correlation Functions: Real Space}

We note that the static density-density correlations are simply
related to the Fourier transforms of the moments $I_0(q).$ Writing the
density correlation function in the form
\begin{equation}
k(r,0) = d \ \delta(r) + d^2 \ C(r) \ ,
\label{eq:defC}
\end{equation}
the dimensionless correlation function $C(r)$ satisfies the relations
$\lim_{r \rightarrow 0 } C(r)\rightarrow - 1 $ and $\lim_{r
\rightarrow \infty }C(r)\rightarrow 0 $. It may be written as
\begin{equation}
C(r) = \frac{1}{d} \int_{- \infty}^{\infty} \frac{ dq}{2 \pi}
\cos(qr)[I_0(q)-1] \ .
\label{eq:defC2}
\end{equation}
The correlation function, in scaled variables, has the representation
\begin{equation}
C(\hat{r} /d)= \int_{-\infty}^{\infty} \; \frac{d \hat{q}} {2}
\cos(\pi \hat{q} \hat{r}) [I_0(\hat{q} k_F) -1] \ .
\label{eq:defC3}
\end{equation}
Explicit expressions are available for the various ensembles from
Mehta \cite{mehtacorrs} by noting that $C(r= \hat{r} /d)$ is nothing
but the two-level cluster function $-Y_2(\hat{r})$, where $\hat{r}$ is
the separation in units of the average interparticle spacing $1/d$.

\subsection{Dynamical Correlations}
\label{sec:dynamical}

We now recapitulate the results from
Ref.\cite{Ben_Lee_Boris_PRL,Ben_Lee_Boris_PRB} for the dynamical
correlation function and present explicit expressions for
$S(q,\omega)$. We note that $S(q,\omega)$ is real and positive;
moreover, it vanishes for $\omega <0$ and it depends only on the
absolute value of $q$.

\subsubsection*{Unitary Ensemble}

We will first examine the simplest case when $\beta=2$ and the system
is non-interacting, which corresponds to the unitary ensemble. It can
be readily shown that
\begin{equation}
k^{u}(r,t) = \frac{d^{2}}{2}
\int_{1}^{\infty}d\lambda_{1}\int_{-1}^{1}d\lambda \exp [-i k_F^{2} t
(\lambda_{1}^{2}-\lambda^{2})] \cos[ k_F r(\lambda_{1}-\lambda)] \ ,
\label{eq:unitary}
\end{equation}
where $\lambda = k/k_{F}, \lambda_{1} = k_{1}/k_{F}$, and $k_{F} = \pi
d$ ($k_{F}$ is the Fermi momentum). Taking the Fourier transform in
both space and time we get
\begin{eqnarray}
S^{u}(q,\omega) & = & \frac{1}{2k_{F}^{2}}
\int_{1}^{\infty}d\lambda_{1}\int_{-1}^{1}d\lambda \
\delta(\lambda_{1}^{2}-\lambda^{2}-\omega/k_{F}^{2}) \
\delta(\lambda_{1}-\lambda-|q|/k_{F}) \nonumber \\ & = &
\frac{1}{4k_{F} |q|} \ \theta(\omega+q^{2}-2k_{F} |q|) \
\theta(\omega-q^{2}+2k_{F} |q|) \ \theta(2k_{F} |q|+q^{2}-\omega) \ .
\label{eq:unitfunc}
\end{eqnarray}
In the Fig.~2 we have plotted the region of support corresponding to
Eq.~(\ref{eq:unitfunc}), which is nothing but the particle-hole
continuum (in this case all excitations are in the form of pairs). The
tridimensional plot of $S^{u}(q,\omega)$ is shown in Fig.~3.

{}From Eq.~(\ref{eq:unitfunc}) we can of course compute all the moments
of $S^{u}(q,\omega)$ exactly. The three moments $I^{u}_{1}$,
$I^{u}_{0}$, and $I^{u}_{-1}$ are plotted in Fig.~4 and we remark that
they are in agreement with the sum rules of
Eqs.~(\ref{eq:f-sumrule},\ref{eq:compresum2}) and the identity
Eq.~(\ref{eq:I0u}).

\subsubsection*{Orthogonal Ensemble}

Secondly, we will consider the orthogonal (attractive) case, when
$\beta=1$. This value of the coupling constant leads to a more
complicated expression for the two-point correlation function; after
Ref.~\cite{Ben_Lee_Boris_PRL}, we have
\begin{eqnarray}
k^{o}(r,t) & = & d^{2} \int_{-1}^{1}d\lambda\int_{1}^{\infty}
d\lambda_{1} \int_{1}^{\infty} d\lambda_{2} \
\frac{(1-\lambda^{2})(\lambda_{1}\lambda_{2}-\lambda)^{2}}
{(\lambda_{1}^{2}+\lambda_{2}^{2}+\lambda^{2}-2\lambda\lambda_{1}
\lambda_{2}-1)^{2}}
\nonumber \\ & \times & \exp[-i k_F^{2}t(2\lambda_{1}^{2}
\lambda_{2}^{2} - \lambda_{1}^{2}-\lambda_{2}^{2}-\lambda^{2}+1)/2]
\cos[ k_F r(\lambda_{1}\lambda_{2}-\lambda)] \ .
\label{eq:orthogonal}
\end{eqnarray}
Taking the Fourier transform of $k^{o}(r,t)$ in space and time yields
\begin{eqnarray}
S^{o}(q,\omega) & = & \frac{2q^{2}}{k_{F}^{4}} \int_{1}^{\infty}
d\lambda_{1} \int_{1}^{\infty} d\lambda_{2} \
\frac{[1-(\lambda_{1}\lambda_{2}-|q|/k_{F})^{2}]}
{(\lambda_{1}^{2}+\lambda_{2}^{2}+q^{2}/k_{F}^{2}-\lambda_{1}^{2}
\lambda_{2}^{2}-1)^{2}} \nonumber \\ & \times &
\delta(\lambda_{1}^{2}+\lambda_{2}^{2}+q^{2}/k_{F}^{2} - 1 -
\lambda_{1}^{2} \lambda_{2}^{2}-2\lambda_{1}\lambda_{2}|q|/k_{F} +
2 \omega/k_{F}^{2}) \nonumber \\
& \times & \theta(\lambda_{1}\lambda_{2}-|q|/k_{F} +1) \
\theta(1-\lambda_{1}\lambda_{2}+|q|/k_{F} ) \ .
\label{eq:orthfunc}
\end{eqnarray}
The region of support of $S^{o}(q,\omega)$ is plotted in Fig.~5. and
it shows a continuum similar to the excitation of a single
particle-hole pair in the non-interacting case. However, as we will
later discuss in detail, the excited states have a more complex
structure. The tridimensional plot of $S^{o}(q,\omega)$ obtained by
numerically integrating Eq.~(\ref{eq:orthfunc}) and using a Gaussian
regularization of the delta function is shown in Fig.~6. The ridge
along $\omega=q(q+k_{F})$ indicates an algebraic (inverse square root)
divergence.

The evaluation of $I_{0}^{o}$ starting from Eq.~(\ref{eq:orthfunc})
has been done analytically by Efetov \cite{Efetov} and it can be
readily generalized for any $I_{n}^{o}$, $n>0$. However, we have only
been able to evaluate numerically the moments with $n<0$. The moments
$I_{1}^{o}$, $I_{0}^{o}$, and $I_{-1}^{o}$ are plotted in Fig.~7 and
they obey exactly the sum rules of Eqs. (\ref{eq:f-sumrule}) and
(\ref{eq:compresum2}), and the identity Eq.~(\ref{eq:I0o}).

\subsubsection*{Symplectic Ensemble}

Finally, we look at the repulsive case, when $\beta=4$. In the context
of RMT this corresponds to the symplectic ensemble. We start with the
correlation function originally obtained in
Ref.~\cite{Ben_Lee_Boris_PRB},
\begin{eqnarray}
k^{s}(r,t) & = & \frac{d^{2}}{2} \int_{1}^{\infty}d\lambda
\int_{-1}^{1} d\lambda_{1} \int_{-1}^{1} d\lambda_{2} \
\frac{(\lambda^{2}-1)(\lambda-\lambda_{1}\lambda_{2})^{2}}{
(\lambda_{1}^{2} + \lambda_{2}^{2}+\lambda^{2}-2\lambda\lambda_{1}
\lambda_{2}-1)^{2}} \nonumber \\ & \times &
\exp[-4i k_F^{2}t(\lambda_{1}^{2}+\lambda_{2}^{2}+\lambda^{2}-2
\lambda_{1}^{2} \lambda_{2}^{2}-1)] \cos [ 2 k_F r(\lambda-\lambda_{1}
\lambda_{2})] \ .
\label{eq:symplectic}
\end{eqnarray}
We take the Fourier transform of $k^{s}(r,t)$ to get
\begin{eqnarray}
S^{s}(q,\omega) & = & \frac{q^{2}}{64k_{F}^{4}} \int_{-1}^{1}
d\lambda_{1} \int_{-1}^{1} d\lambda_{2} \
\frac{[(\lambda_{1}\lambda_{2}+|q|/2k_{F}
)^{2}-1]}{(\lambda_{1}^{2}+\lambda_{2}^{2}+q^{2}/4k_{F}^{2}-
\lambda_{1}^{2} \lambda_{2}^{2}-1)^{2}} \nonumber \\ & \times &
\delta(\lambda_{1}^{2}+\lambda_{2}^{2}+q^{2}/4k_{F}^{2} - 1 -
\lambda_{1}^{2} \lambda_{2}^{2} + \lambda_{1}\lambda_{2}|q|/k_{F}
- \omega/4k_{F}^{2}) \nonumber \\ & \times & \theta(\lambda_{1}
\lambda_{2}+|q|/2k_{F} - 1) \ .
\label{eq:sympfunc}
\end{eqnarray}
The region of support of $S^{s}(q,\omega)$ is shown in Fig.~8. The
continuum reaches $\omega=0$ not only at $q=0$ and $2k_{F}$, but at
$q=4k_{F}$ as well. The figure is similar to the one obtained when we
excite two particle-hole pairs in the non-interacting case; in fact,
the real structure of the excited states is more complicated than this
simple picture, as we will later demonstrate. In Fig.~9 we show the
tridimensional plot of $S^{s}(q,\omega)$ with a Gaussian
regularization and a numerical integration of Eq.~(\ref{eq:sympfunc}).
The ridges along $\omega=2q(2k_{F}-q)$ and $\omega=q^{2}-4k_{F}^{2}$
indicate an algebraic (inverse square root) divergence.

As for the orthogonal ensemble, the analytical evaluation of
$I_{n}^{s}$ can be done for $n>0$ by generalizing the method of
Ref.~\cite{Efetov} for $n=0$. For $n<0$ we have only been able to
proceed with a numerical evaluation. In Fig.~10 we have plotted the
moments $I_{1}^{s}$, $I_{0}^{s}$, and $I_{-1}^{s}$. It is simple to
check that the moments obtained in this way also agree with
Eqs.~(\ref{eq:f-sumrule},\ref{eq:compresum2},\ref{eq:I0s}).

\subsection{Spectrum Saturation in the Hydrodynamic Limit}

An important property of the Sutherland model is the saturation at
$q\rightarrow 0$. As we have pointed out before in
Eqs.~(\ref{eq:omegaF_limit},\ref{eq:omegaH}), both the Feynman and
hydrodynamical spectra tend to the same value in this limit. In fact,
this is also true for any characteristic frequency defined as the
ratio of any two moments. We shall prove it for the orthogonal
ensemble: In Eq.~(\ref{eq:orthfunc}) we change variables to $$
\lambda_{1,2}=1+x_{1,2}, $$ which yields
\begin{eqnarray}
I_{n}^{o}(q) & = & \frac{2q^{2}}{k_{F}^{4}}
\left(\frac{k_{F}^{2}}{2}\right)^{n+1} \int_{0}^{\infty} \ dx_{1}
\int_{0}^{\infty} dx_{2} \
\frac{[1-(1+x_{+}+x_{1}x_{2}-q/k_{F})^{2}]}{[q^{2}/k_{F}^{2} -
x_{1} x_{2} (4+x_{1}x_{2}+2x_{+})]^{2}} \nonumber \\
& \times & \left[ -\frac{q^{2}}{k_{F}^{2}} +
\frac{2q(1+x_{+}+x_{1}x_{2})}{k_{F}} + x_{1}x_{2}(4 + x_{1}x_{2} +
2x_{+})\right]^{n} \nonumber \\ & \times &
\theta(2+x_{+}+x_{1}x_{2}-q/k_{F}) \ \theta(
q/k_{F}-x_{+}-x_{1}x_{2})\ ,
\label{eq:saturo1}
\end{eqnarray}
where $x_{\pm}=x_{1}\pm x_{2}$. This last expression can be much
simplified in the limit $q\rightarrow 0$; it becomes
\begin{eqnarray}
I_{n}^{o}(q) & \stackrel{q\rightarrow 0}{\longrightarrow} &
\frac{2(qk_{F})^{n+2}}{k_{F}^{4}}
\int_{0}^{q/k_{F}} \ dx_{1} \int_{0}^{q/k_{F}-x_{1}} dx_{2} \
\frac{(q/k_{F}-x_{+})}{(q^{2}/k_{F}^{2}-x_{+}^{2}+x_{-}^{2})^{2}} \ .
\label{eq:saturo2}
\end{eqnarray}
Changing the integration variables to $x_{\pm}$ and performing the
double integral we obtain
\begin{eqnarray}
I_{n}^{o}(q) & \stackrel{q\rightarrow 0}{\longrightarrow} &
\frac{1}{k_{F}^{2}}(q k_{F})^{n+1} \ .
\label{eq:saturo3}
\end{eqnarray}
As a result,
\begin{eqnarray}
\left[\frac{I_{n+m}^{o}(q)}{I_{n}^{o}(q)}\right]^{1/m}_{q
\rightarrow 0} \rightarrow q k_{F} \ .
\label{eq:saturo4}
\end{eqnarray}
The proof for the symplectic and unitary ensembles is quite analogous;
one obtains for the three values of $\beta$
\begin{equation}
I_{n}(q) \stackrel{q\rightarrow 0}{\longrightarrow}
\frac{1}{\beta^{2}k_{F}^{2}}(\beta q k_{F})^{n+1}
\label{eq:saturs1}
\end{equation}
This yields, for any integers $n$ and $m$,
\begin{equation}
\left[\frac{I_{n+m}(q)}{I_{n}(q)}\right]^{1/m}_{q \rightarrow 0}
\rightarrow \beta qk_{F} \ .
\label{eq:saturs2}
\end{equation}

The saturation property can also be visualized in Figs.~2, 5, and
8: the fact that the lower and upper branches of parabola have the
same linear term as $q\rightarrow 0$ implies that all characteristic
frequencies must have the same asymptotics.

\section{ Asymptotic Bethe's Ansatz}
\label{sec:ABA}

We now summarize the results of the asymptotic Bethe's Ansatz, which
gives an explicit expression for the ``particle-hole'' like
excitations underlying the system. The {\it complete} excitation
spectrum for the $1/r^2$ model can be described in remarkably simple
terms as follows: The total energy of a state of the system is
expressible as
\begin{equation}
E = \sum_{n} \ p_{n}^{2} \ ,
\label{eq:energBethe}
\end{equation}
with the ``pseudo-momenta'' $p_n$ satisfying the equation,
\begin{equation}
p_{n} = k_{n} + \frac{\pi(\beta-2)}{2L}\sum_{n\neq
m}\mbox{sign}{(k_{n}-k_{m})} \ .
\label{eq:pseudomom}
\end{equation}
The total momentum of the state is
\begin{equation}
P = \sum_{n} \ p_{n} = \sum_n \ k_n \ .
\label{eq:momentBethe}
\end{equation}
The bare momenta are given by $k_{n} = 2\pi J_{n}/L$, where the
$J_{n}$'s are fermionic quantum numbers $J_1 < J_2 <J_3 \ldots < J_N$.
Note that at $\beta=2$ the interaction is turned off and we recover
the free-fermion results. The important point is that the totality of
states for the N particle sector is obtained by allowing the integers
$J_n$ to take on all values consistent with Fermi statistics, not only
for $\beta=2$, but for {\it all} $\beta \in [1, +\infty]$. The
summation in Eq.~(\ref{eq:pseudomom}) is trivial to carry out and we
find
\begin{equation}
p_n= k_n + \frac{(\beta-2)\pi}{L} \left(n- \frac{N+1}{2}\right) \ .
\label{eq:p_n}
\end{equation}
We can now select an arbitrary state of the system by specifying that
states $\{k_1, k_2, \ldots \}$ are occupied, i.e. by introducing the
fermionic occupation numbers $n(k_j)=0,1$, such that
\begin{eqnarray}
E= \sum_{n} \varepsilon(k_{n})n(k_{n}) + \sum_{n\neq m}
v(k_{n}-k_{m})n(k_{n})n(k_{m}) +
\left[\frac{\pi(\beta-2)}{2}\right]^{2} \ ,
\label{eq:energquasi}
\end{eqnarray}
with $\varepsilon(k)=k^{2}$ and $v(k)=\pi(\beta-2)|k|/2L$. For future
reference, the ground state is represented by $n_0(k_{n})=1$ for
$|k_{n}|<k_{F}$ and $n_0(k_{n})=0$ otherwise, where $k_F= \pi d$. We
remark that the above expression of the energy takes the form of a
renormalized Hartree-Fock theory; a Hartree-Fock energy expectation
value of the interacting Hamiltonian in a determinantal state $\prod
c^{\dagger}_{k_j} |0\rangle$ leads to precisely this type of
expression. Note that the Fourier transform of the two-body
interaction can be deduced from the expansion $$(\frac{\pi}{L})^2
\frac{\beta (\beta-2)}{\sin^2(\pi r/L)} = - \beta(\beta-2)
\frac{\pi}{L} \sum_q |q| \exp(i q r) \ .$$ Therefore,
Eq.~(\ref{eq:energquasi}) states that a Hartree-Fock expression with a
renormalization of the coupling constant $\beta (\beta-2)
\longrightarrow 2 (\beta-2)$ leads to the {\it exact spectrum} of the
model.

We now consider the excitation spectrum near the ground state, wherein
we excite a particle-hole pair in the free Fermi system and ask what
the energy of the interacting system is by including the Hartree-Fock
back flow term. From this point onwards we measure all momenta in
units of $k_F$ and energies in units of the Fermi energy. Let us
suppose that one of the particles described by
Eq.~(\ref{eq:energquasi}) has initially a momentum $k$, with $|k|<1 $;
we promote it to some state labeled by $k+q$, with $|k+q|> 1$. The
energy cost in units of the Fermi energy is equal to
\begin{eqnarray}
\triangle(q,k) & = & \varepsilon(k+q) - \varepsilon(k) + 2
\sum_{|k'|<1} [v(k+q-k')-v(k-k')] \nonumber \\
& = & q^{2}+ 2 k q + \frac{(\beta-2)}{2}(2|k+q|-k^{2}-1) \ ,
\label{eq:energydiff}
\end{eqnarray}
and the momentum of this state is simply $q$. This implies that we can
associate a generalized energy corresponding to a particle
$\varepsilon_{>}(k)$ (i.e. $|k| > 1$) and a hole $\varepsilon_{<}(k)$
(i.e. $|k| < 1$):
\begin{eqnarray}
\varepsilon_{>}(k)&=&k^2 + (\beta-2) |k| \nonumber \\
\varepsilon_{<}(k)& = &\frac{\beta}{2} k^2 +\left( \frac{\beta}{2}
-1 \right) \ ,
\label{eq:quasienergies}
\end{eqnarray}
such that
\begin{equation}
\triangle(q,k) = \varepsilon_{>}(k+q)-\varepsilon_{<}(k) \ .
\label{eq:triang}
\end{equation}
Note that $\varepsilon_{>}(k)$ and $\varepsilon_{<}(k)$ are continuous
and have continuous derivatives across the Fermi surface. These
expressions for the particle and hole energies look different from the
results of Sutherland \cite{Sutherland_course} but their equivalence
may be readily checked. We have chosen to label our states by the
``bare'' momenta ``$k_n$'', while Ref.\cite{Sutherland_course} works
with the pseudomomenta ``$p_n$''; of course these are in one to one
correspondence and so the choice is a matter of convenience.

We will introduce in the usual way, particle operators
$A^{\dagger}(k)$ and hole operators $B^{\dagger}(k)$ with the
convention that the momenta corresponding to these are constrained by
$|k| > k_F$ for particles and $|k| \leq k_F$ for holes, with
excitation energies
\begin{eqnarray}
E_A(k) & = & \varepsilon_>(k)- \mu \nonumber \\
E_B(k) & = & \mu - \varepsilon_<(-k) \ ,
\label{eq:quasiph}
\end{eqnarray}
where $\mu \equiv \varepsilon_>(k_F)$ is the ``chemical potential''.
The quasi-particle quasi-hole excitation created by the operator $
A^{\dagger}(k+q) B^{\dagger}(-k)$ then has energy $E_A(k+q)+ E_B(-k)$,
which of course is equal to $\triangle(k,q)$. Having introduced the
underlying fermionic quasi-particles quasi-holes through
Eqs.~(\ref{eq:quasienergies}), we would like to see if the excitations
generated by the bare density fluctuation operator $\rho_q$ can be
expressed in terms of the latter. One of our objectives then, is to
express the excitations of the system probed by the bare density
fluctuation operator $\rho_q$ in terms of the quasi-particle
quasi-hole operators. Recall that in Landau's Fermi Liquid Theory
\cite{Landau} one expresses the bare particles $c(k)$ in a series
involving quasi-particles and quasi-holes of the form
\begin{equation}
c(k)= \sqrt{z_k} \ B^{\dagger}(-k) + \sum_{(p,l)} M[k,p,l] \
B^{\dagger}(p) B^{\dagger}(l) A^\dagger(-k-p-l) + \ldots ,
\label{eq:landau}
\end{equation}
where $ | k | \leq k_F$, and a similar expansion for particles, where
$z_k$ is the quasi-particle residue. The density fluctuation operator
$\rho_q = \sum_k c^\dagger(k+q)c(k)$ then has a development in terms
of $1, 2, 3, \ldots$ pairs of (quasi) particle-hole excitations. In
one dimension, we expect $z_k$ to vanish for arbitrary non-zero
interactions, and hence the particle-hole series is expected to be
such that the single pair should not appear. The expansions are
somewhat non-unique, in view of the fact that we can add an arbitrary
number of ``zero energy'' and ``zero-momentum'' particle-hole
excitations to any given scheme.

\section{Quasi-Particle Content of the Structure Function}
\label{sec:content}

For the unitary case there is no interaction and consequently
quasi-particles and quasi-holes are the same as particles and holes:
Eq.~(\ref{eq:energydiff}) at $\beta=2$ exactly describes the spectrum
of Fig.~2. On the other hand, for the orthogonal and symplectic cases
the simple creation of quasi particle-hole pairs cannot account for
the whole excitation spectrum. In order to see that in general
($\beta\neq 2$), we begin by considering the spectrum
($q\times\omega$) for a particle-hole excitation (Eq.
(\ref{eq:energydiff})), which is the familiar pair spectrum
renormalized by the interaction. For a fixed $q$, the maximum value of
$\omega$ occurs when $k=1$: $\omega_{max}=q^{2}+\beta|q|$. The minimum
value of $\omega$ depends on $q$: for $|q|<2 $ it occurs when $k=1-|q|
$; for $|q|>2 $ it occurs when $k=-1$. This results in
$\omega_{min}=\beta |q|(2-|q|)/2$ for $q<2$, and $\omega_{min}=(|q|-2
)[|q|+(\beta-2)]$, for $q>2$. The curves bound a continuum which does
not agree with the hashed regions of either Figs.~5 or 8.

We can also promote two quasi-particles from the Fermi sea, instead of
just one. The result is that the upper limit for $\omega$ is then
given by $|q|(|q|+\beta )$ and the lower limits are $\beta
|q|(2-|q|)/2$, $\beta(|q|-2 )(4 -|q|)/2$, and $(|q|-4
)[|q|+2(\beta-2)]/2$ (for the intervals $|q|<2$, $2<|q|<4$, and
$|q|>4$, respectively). Again, we note that the continuum bound by
these curves is not equal to the hashed region in Fig.~8.

Below we recast the expressions for the correlation functions in the
various ensembles in terms of new variables, in order to reveal their
exact quasi-particle content.

\subsection*{Orthogonal Ensemble: Change of Variables}

We now turn to the expression Eq.~(\ref{eq:orthfunc}) of the structure
function. Firstly we change variables and introduce
\begin{eqnarray}
u & =& \lambda_1 \lambda_2  \nonumber \\
z & =& \lambda_1 + \lambda_2 \ .
\label{eq:newvars}
\end{eqnarray}
With this change of variables, we  find
\begin{equation}
S^{o}(q,\omega) = 2 q^2 \int_{max \{1,q-1 \}}^{(1+q)} \ du \ \int_{2
\sqrt{u}}^{(1+u)} \ dz \frac{1}{\sqrt{z^2- 4 u}} \ [1-(u-q)^2]
\delta(E) / D \ ,
\label{eq:s_o1}
\end{equation}
where
\begin{eqnarray}
\sqrt{D} & = & z^2 +q^2- (1+u)^2 \nonumber \\
E & = & \sqrt{D} + 2 \omega - 2 u q \ .
\label{eq:arg_o}
\end{eqnarray}

In the expressions above and hereafter in this section we will set $q
> 0$ without loss of generality. We change the integration variable
$u$ by defining $k=u-q$, in terms of which the $k$ integration is
restricted to $max\{1-q,-1\} \leq k \leq 1$, and is immediately
recognizable as the momentum of a hole restricted to the Fermi surface
with $|k+q|$ restricted to be a particle. The $z$ integration can be
conveniently transformed by introducing a ``rapidity'' variable
\begin{equation}
z= 2 \sqrt{(k+q)} \cosh \theta \ ,
\label{eq:rapid_o}
\end{equation}
and recalling that for the orthogonal case $\varepsilon_>(k)= k^2
-|k|$ and $\varepsilon_<(k)=(k^2-1)/2$ (see
Eq.(\ref{eq:quasienergies})), so that
\begin{eqnarray}
S^{o}(q,\omega) & = & \frac{q^{2}}{2} \int_{ \buildrel {|k| \leq 1}
\over {k+q > 1}} \ dk \ \frac{[-\varepsilon_{<}(k)]}{[\omega -
q(k+q)]^{2}} \int_{0}^{\ln\sqrt{(k+q)}} \ d\theta \nonumber \\ &
\times & \delta( \varepsilon_{>}(k+q) - \varepsilon_{<}(k) -
2(k+q)\sinh^2\theta -\omega ) \ .
\label{eq:s_o2}
\end{eqnarray}
The $\theta$ integration can be done simply, and gives the final result
\begin{equation}
S^{o}(q,\omega)= \frac{q^{2}}{4} \int_{ \buildrel {|k| \leq 1} \over
{k+q > 1}} \ dk \ \frac{[-\varepsilon_{<}(k)]}{[\omega - q(k+q)]^{2}}
\frac{\theta(\omega- \triangle(k,q)+(k+q-1)^{2}/2) \
\theta(\triangle(k,q)-\omega)}{\sqrt{\triangle(k,q)-\omega} \
\sqrt{\triangle(k,q)+2(k+q)-\omega }}.
\end{equation}

\subsubsection*{Bethe Quasi Particle-Hole Content: Orthogonal Ensemble}

We can rewrite the energy conserving delta function in
Eq.~(\ref{eq:s_o2}) as $\delta(
\varepsilon_{>}^{(o)}(k+q|\theta)-\varepsilon_{<}(k) - \omega)$, where
$\varepsilon_{>}^{(o)}(k| \theta) = \varepsilon_>(k) - 2 k \sinh^2
\theta$. One possible picture suggested then has the excited state
particle possessing a ``hidden'' gauge variable $\theta$, which lies
in a limited range, as a hole would, endowed with energy but
possessing no ``physical momentum''. The particle say for $k > 1$ has
an energy ${\varepsilon_{>}^{(o)}}(k)$ which lies between
$\frac{k^2-1}{2}$ and $k^2 - k$. We can also view the excited particle
state as a combination of a particle and particle-hole pair, as
follows. We write a schematic development for $k \geq 1$
\begin{equation}
c^{\dagger}(k) \sim \sum_{\frac{1+k}{2} \leq p \leq k} A^{\dagger}(p)
A^{\dagger}(k-p+1) B^{\dagger}(-1) \ ,
\label{eq:c_dag}
\end{equation}
The excitation energy of this complex is readily seen from
Eqs.~(\ref{eq:quasienergies},\ref{eq:quasiph}) to be $E_A(p) +
E_A(k-p+1) + E_{B}(-1)$, with $\frac{1+k}{2} \leq p \leq k$. The sum
of these three terms reproduces the variation in
$\varepsilon_{>}^{(o)}(k| \theta)$ implied by the rapidity variable.
The density fluctuation $\rho_{q}$ is then seen to be formally a two
quasi particle-hole object: writing $c(k) \sim B^{\dagger}(-k)$, we
have
\begin{equation}
c^{\dagger}(k) c(k-q) \sim \sum_{\frac{1+k}{2} \leq p \leq k}
A^{\dagger}(p) A^{\dagger}(k-p+1) B^{\dagger}(-1) B^\dagger(q-k) \ ,
\label{eq:bubble_orth}
\end{equation}
where it is understood here and elsewhere that when two momenta
coincide (as they would in say $[B^{\dagger}(-1)]^2$), then these
should be separated by the smallest non-zero wave number. The above
scheme for the density fluctuation operator $c^{\dagger}(k)c(k-q)$ is
indicated in Fig.~11. We may therefore regard the density fluctuation
as being built up from a particular set of (non-interacting) pair
states consisting of annihilating two particles at momenta $k-q$ and
$1$, and creating a pair with total momentum $k+1$, distributed over
all possible relative momenta with appropriate form factors.

\subsection*{Symplectic Ensemble: Change of Variables}

We recall for the symplectic case, the Bethe energies
$\varepsilon_{>}(p)= p^2 +2 |p|$ and $\varepsilon_{<}(p)=2 p^2 + 1$
(Eq. \ref{eq:quasienergies})). We now rewrite Eq.~(\ref{eq:sympfunc})
using the same variables as in the previous case
(Eq.~(\ref{eq:newvars})). We find the result breaks up naturally into
two pieces $S_a$ and $S_b$, with the second piece $S_b$ only arising
for $q>2$:
\begin{equation}
S^s(q,\omega) = S_a( q,\omega) + \theta(q-2) \ S_b(q,\omega ) \ ,
\label{eq:twoparts}
\end{equation}
with
\begin{equation}
S_a(q,\omega) = \frac{q^2}{16} \int_{ max(0,1-q/2)}^1 \ du \
\int_{2\sqrt{u}}^{(1+u)} \frac{dz}{ \sqrt{z^2- 4u}} \ \delta(E_a)
\frac{N_a}{D_a} \ ,
\label{eq:sympinter_a}
\end{equation}
and
\begin{equation}
S_b(q,\omega) = \frac{q^2}{16} \int_{ max\{-1,1-q/2\}}^{0} \ du \
\int_{0}^{(1+u)} \ \frac{dz}{\sqrt{z^2-4u}} \ \delta(E_b)
\frac{N_b}{D_b} \ ,
\label{eq:sympinter_b}
\end{equation}
where $E$, $N$ and $D$ are appropriately defined (see below). We write
$u=(1+k)/2$ in $S_a$ and $u=(l-1)/2$ in $S_b$, in terms of which
Eqs.~(\ref{eq:sympinter_a}) and (\ref{eq:sympinter_b}) take a more
natural form
\begin{equation}
S_a(q,\omega) = \frac{q^2}{32} \int \ dk \ n_{0}(k) [1- n_{0}(k+q)] \
\int_{\sqrt{2(1+k)}}^{(3+k)/2} \frac{dz}{ \sqrt{z^2- 2(1+k)}} \
\delta(E_a) \frac{N_a}{D_a} \ ,
\label{eq:symp2_a}
\end{equation}
and
\begin{equation}
S_b(q,\omega) = \frac{q^2}{32} \int \ dl \ n_{0}(l) [1- n_{0}(l+q-2)]
\ \int_{0}^{(1+l)/2} \frac{dz}{\sqrt{z^2 +2(1-l)}} \ \delta(E_b)
\frac{N_b}{D_b} \ . \label{eq:symp2_b}
\end{equation}

In Eq.~(\ref{eq:symp2_a}) we further introduce the rapidity variable
$\theta$ through $z=8(1+k)\mbox{sinh}^2\theta$, so as to eliminate the
square root in the integrand. The result can be written compactly as
follows
\begin{eqnarray}
S_a(q,\omega) & = & \frac{q^2}{2} \int_{ \buildrel {|k| \leq 1} \over
{k+q > 1}} \ dk \frac{[\varepsilon_{>}(k+q) -3]}{[\omega - 2 q
(1+k)]^{2}} \int_{0}^{\ln\sqrt{2/(1+k)}} d\theta \nonumber \\ & \times
& \delta(\triangle(k,q) + \ 8(1+k) \sinh^2\theta-\omega) \ .
\label{eq:symp3_a}
\end{eqnarray}
We can perform explicitly the rapidity integrals and find the final
result
\begin{eqnarray}
S_a(q,\omega) & = & \frac{q^2}{4} \int_{ \buildrel {|k| \leq 1} \over
{k+q > 1}} \ dk \frac{[\varepsilon_{>}(k+q) -3]}{[\omega -
\triangle(k,q)-(k-1)^2+q^2]^2} \nonumber \\ & \times &
\frac{\theta(\omega- \triangle(k,q)) \ \theta(\triangle(k,q) +(k-1)^2-
\omega)}{\sqrt{\omega- \triangle(k,q)} \sqrt{\omega- \triangle(k,q)
+8(1+k)}} \ .
\label{eq:symp4_a}
\end{eqnarray}

We next turn to the other piece for $q>2$. With $\alpha \equiv (q-2)$
and introducing the rapidity variable $\phi$ through $z=8(1-l)
\mbox{sinh}^2(\phi)$, Eq.~(\ref{eq:symp2_b}) is expressible in the
form
\begin{eqnarray}
S_b(q,\omega) & = & \frac{q^2}{2} \int_{ \buildrel {|l| \leq 1} \over
{l + \alpha > 1}} \ dl \frac{[\varepsilon_{>}(l+ \alpha) -3]}{[\omega
- 2 q (l-1)]^2} \ \int_{0}^{\ln\sqrt{2/(1-l)}} d\phi \nonumber \\ &
\times & \delta(\triangle(l, \alpha) + \, 8(1-l) \sinh^2 \phi -\omega)
\ .
\label{eq:symp3_b}
\end{eqnarray}
Performing the rapidity integration we find
\begin{eqnarray}
S_b(q,\omega) & = & \frac{q^2}{4} \int_{ \buildrel {|l| \leq 1} \over
{l +\alpha > 1}} \ dl \ \frac{[\varepsilon_{>}(l+ \alpha) -3]}{[\omega
- \triangle(l,\alpha)-(l+1)^2+q^2]^2} \nonumber \\ & \times &
\frac{\theta(\omega - \triangle(l,\alpha )) \
\theta(\triangle(l,\alpha) +(l+1)^2 -\omega)}{\sqrt{\omega -
\triangle(l, \alpha)} \ \sqrt{\omega - \triangle(l, \alpha) +8(1-l)}}
\ .
\label{eq:symp4_b}
\end{eqnarray}

\subsubsection*{Bethe Quasi Particle-Hole Content: Symplectic Ensemble}

The energy conserving delta function in Eq.~(\ref{eq:symp3_a}) can be
rewritten using an effective energy variable
$\varepsilon_<^{a}(k|\theta) = \varepsilon_<(k) - \, 8(1+k) \sinh^2
\theta$, which implies $\omega= \varepsilon_>(k+q)
-\varepsilon_<^a(k|\theta)$. The energy $\varepsilon_<^a(k|\theta)$
varies between the limits $2 k^2 +1 $ (i.e. $\varepsilon_<(k)$) and
$k^2 +2 k$ (i.e. $\varepsilon_<(k) -(k-1)^2$). It is thus evident that
we may interpret $\varepsilon_<^a(k|\theta)$ as an effective hole, i.e
a composite object. One possible way to decompose it is to write
schematically for the bare annihilation operator a representation as a
quasi-hole plus a quasi particle-hole pair:
\begin{equation}
c(k) \sim \sum_{k\leq p \leq \frac{k+1}{2}} B^{\dagger}(p-k-1)
B^{\dagger}(-p) A^{\dagger}(1) \ .
\label{eq:c_a}
\end{equation}
The restriction on the range of $p$ is such that we avoid double
counting the pair and have a natural ordering of the two quasi-holes.
The energy of the effective hole is then $E_B(-p) + E_B(p-k-1) +
E_{A}(1)$, with the constraint $k \leq p \leq \frac{k+1}{2}$, which
from Eq.~(\ref{eq:quasienergies},\ref{eq:quasiph}) reproduces the
range required by the rapidity variation. Therefore, the operator
$\rho_{q}$ is seen to be formally a two quasi particle-hole object,
\begin{equation}
c^\dagger(k+q) c(k) \sim \sum_{k\leq p \leq \frac{k+1}{2}}
B^{\dagger}(p-k-1) B^{\dagger}(-p) A^{\dagger}(1) A^\dagger(k+q) \ .
\label{eq:bubble_a}
\end{equation}
This scheme for the density fluctuation is illustrated in Fig.~11.

In the second piece of $S^{s}$ (Eq.~(\ref{eq:symp3_b})), once again the
energy conserving delta function can be rewritten introducing an
effective energy variable for the hole $\varepsilon_<^{b}(l |\phi) =
\varepsilon_<(l) - \, 8(1-l) \sinh^2\phi$, which implies $\omega=
\varepsilon_>(l + \alpha)-\varepsilon_<^b(l| \phi)$. The effective hole
energy $ \varepsilon_<^b(l| \phi)$ varies between $\varepsilon_<(l)$
and $\varepsilon_<(l)- (1+l)^2$, corresponding to a predominantly left
moving object, and may be decomposed again into a hole and a
particle-hole pair. Schematically we have $$c(l) \sim
\sum_{\frac{l-1}{2} \leq p \leq l} B^{\dagger}(-p) A^\dagger(-1)
B^{\dagger}(1+p-l) \ . $$ Using the quasi-energies
Eqs.~(\ref{eq:quasienergies},\ref{eq:quasiph}) this complex has energy
$E_B(p) + E_B(1+p-l) + E_{A}(-1)$, with the physical constraint
$\frac{l-1}{2} \leq p \leq l$, which reproduces the range implied by
the variation of the rapidity. Owing to momentum conservation, we must
regard the creation operator $c^\dagger(l+q)$ as $A^\dagger(l+q-2)$
times a particle-hole pair with energy zero and momentum $2$, i.e.
$A^\dagger(1)B^\dagger(1)$. We may eliminate a `zero pair'
$A^{\dagger}(-1)B^{\dagger}(1)$ and thus obtain the scheme for the
density fluctuation operator $\rho_{q}$,
\begin{equation}
c^{\dagger}(l+q)c(l) \sim \sum_{\frac{l-1}{2} \leq p \leq l}
A^{\dagger}(l+q-2) A^{\dagger}(1) B^{\dagger}(-p) B^{\dagger}(1+p-l) \
{}.
\label{eq:bubble_b}
\end{equation}
The term $S_b$ then evidently may be regarded as a two quasi
particle-hole object, and is illustrated in Fig.~11.

Summarizing, in process ($a$), we may regard the density fluctuation
as being built up from a particular set of (non-interacting) pair
states consisting of creating two particles at momenta $1$ and $k+q$,
and destroying a pair with total momentum $k+1$, distributed over all
possible relative momenta with appropriate form factors. Likewise, in
process ($b$), we may regard the density fluctuation as being built up
from a particular set of (non-interacting) pair states consisting of
creating two particles at momenta $1$ and $l+q-2$, and destroying a
pair with total momentum $l-1$, distributed over all possible relative
momenta with appropriate form factors.

\section{Spin 1/2 Heisenberg systems}
\label{sec:discrete}

In this section we will study the moments $I_n$ in two standard spin
1/2 Heisenberg antiferromagnetic systems, the $1/r^2$ system and the
Bethe chain. The Heisenberg spin chain model with a $1/r^{2}$
interaction was introduced by Haldane \cite{Haldane} and Shastry
\cite{Shastry88}. It is defined by the Hamiltonian
\begin{equation} H = J \phi ^2 \sum_{i<j}
\frac{\vec{S}_{i}\cdot\vec{S}_{j}}{\sin^{2}[\phi(r_{i}-r_{j})]} \ ,
\label{eq:hamilt_discrete}
\end{equation}
where $\phi= \frac{ \pi}{L}$, $r_{i}=0,1,... ,L-1$ ($L$ integer), and
the spins are $1/2$. In this case the natural operators one can use to
introduce a dynamical structure function $S(Q,\omega)$ are the
$\hat{S_{i}^{z}}$: we define a ``charge operator''
$\hat{\rho_{i}}=(\hat{S_{i}^{z}}+1/2)$ and use (\ref{eq:sqw}) to write
\begin{equation}
S^{d}(Q,\omega) \equiv \frac{1}{N}\sum_{\nu\neq 0}
|\langle\nu|\hat{S}_{Q}|0\rangle|^{2}\delta(\omega-E_{\nu}+E_{0}) \ ,
\label{eq:sqw_disc}
\end{equation}
with $E_{\nu}$ as the energy
eigenvalues of the Hamiltonian and the lattice Fourier transform
\begin{equation}
\hat{S}_{Q}^{z} = \sum_{j=1}^{L} \hat{S}_{j}^{z} e^{-iQr_{j}} \ ,
\label{eq:spinop}
\end{equation}
where $Q$ is the lattice momentum $Q= (2\pi /L)\times \mbox{integer}$.
$N$ is the number of spin deviations or the number of hard-core
bosons. We set $N= \hat{d} L$, so that $\hat{d}=1/2$ for half filling.
The Fermi momentum is then $k_F= \pi \hat{d} $, and we will scale $Q=
\pi \hat{d} \hat{Q}$ in order to compare with the continuum model
results.

We cannot calculate $S^{d}(Q,\omega)$ directly, but we know some
moments of this distribution, just as we did for the continuum model.
As in the continuum, we restrict ourselves to $n=0,1$, and $-1$.

\subsection{Static Correlation Functions and the Zeroth Moment}

We begin with the zeroth moment, or the static structure factor
\begin{equation}
I^{d}_{0}(Q) = \int_{0}^{\infty} S^d(Q,\omega) \; d\omega .
\end{equation}
There is a remarkable theorem by Mehta and Mehta \cite{2Mehtas}
stating that the static correlation function is identical to that of
the repulsive ($\beta=4$) continuum model in real space. This object
was also calculated independently for half filling in
Ref.\cite{Vollhardt}, and the result (in Q space) is for the half
filled case:
\begin{equation}
I_{0}^{d}(Q) = - \frac{1}{2 } \ln \left( 1-\frac{|Q|}{\pi } \right) \ ,
\label{eq:I0d}
\end{equation}
in the scheme where we restrict $ |Q| \leq \pi$.

The correlation function $I_0$ is also available from Mehta and
Mehta\cite{2Mehtas}, for {\it arbitrary} densities $\hat{d} \leq
\frac{1}{2}$. The density-density correlator can be written as
\begin{equation}
\langle\hat{\rho}_{0} \hat{\rho}_{r}\rangle = \hat{d} \delta_{r,0}
+ (1- \delta_{r,0}) \; \hat{d}^2 \left[1 + D(r) \right],
\label{defD}
\end{equation}
in a manner similar to Eq.(\ref{eq:defC}). The function $D$ is given
\cite{2Mehtas} explicitly in the thermodynamic limit as
\begin{equation}
D(r)= - \left[\frac{\sin( 2 \pi r \hat{d})}{2 \pi r
\hat{d}}\right]^{2} + \left(\int_{0}^{2\pi r \hat{d}} dt \frac{\sin
t}{t} \right) \ \frac{(2 \pi r \hat{d}) \cos(2 \pi r \hat{d}) - \sin(2
\pi r \hat{d})}{(2 \pi r \hat{d})^2} \ .
\end{equation}
Using the relation
$\hat{S}_i^z=\hat{\rho}_i- \frac{1}{2}$, we find
\begin{equation}
\langle \hat{S}^{z}_{0} \hat{S}^{z}_{r}\rangle= \frac{1}{4} + \hat{d}
(1 - \hat{d}) (\delta_{r,0}-1) + (1- \delta_{r,0})\; \hat{d}^2 \;
D(r).
\end{equation}
Inverting the Fourier series, we have
\begin{equation}
I^{d}_{0}(Q)= 1 + \hat{d} \; \sum_{r} \exp( i Q r) \; D(r).
\end{equation}
We may convert the sum over $r$ to an integral, remembering that $Q$
and $Q + 2 \pi \times \mbox{integer}$ are equivalent. We will work in
the reduced zone scheme $|Q| \leq \pi$, for which two cases may be
distinguished, case (A) $\hat{d} \leq \frac{1}{4}$ and case (B)
$\frac{1}{4} < \hat{d} \leq \frac{1}{2}$. The correlations are given
for $Q \geq 0$, and may be obtained for negative $Q$ by using the
evenness in $Q$ of $I_0$. For case (A) $\hat{d} \leq \frac{1}{4}$ we
find
\begin{eqnarray}
I^{d}_{0}(Q)& = & 1 + \; \theta(4 \pi \hat{d} - Q) A(Q) \\
A(Q) & = & \frac{Q}{4 \pi \hat{d}} - 1 - \frac{Q}{8 \pi \hat{d}}
\ln\left| 1- \frac{Q}{2 \pi \hat{d}}\right| \ ,
\label{eq:mehtaA}
\end{eqnarray}
and for case (B)  $\frac{1}{4} < \hat{d} \leq \frac{1}{2}$ we find
\begin{eqnarray}
I^{d}_{0}(Q)= 1 + \; A(Q) + \ \theta(Q- 2\pi + 4 \pi \hat{d}) \ A(2
\pi - Q).
\label{eq:mehtaB}
\end{eqnarray}
It can be checked for $\hat{d}=1/2$ that Eq.~(\ref{eq:mehtaB}) is
identical with Eq.~(\ref{eq:I0d}). For $\hat{d} \leq 1/4$, the
expression in Eq.~(\ref{eq:mehtaA}) is identical to the symplectic
case Eq.~(\ref{eq:I0s}), apart from a scale factor of $\hat{d}$.
Eq.~(\ref{eq:mehtaB}) is in fact nothing but the Umklapp reduction of
the continuation of Eq.~(\ref{eq:mehtaA}), i.e. $Q$ is allowed to
extend up to $ 4 \hat{d} \pi$, and the part beyond $\pi$ is declared
to belong to $Q- 2\pi$, after subtracting unity from the structure
function.

The correlation function, in scaled variables, is given by (compare
Eq.~(\ref{eq:defC3}))
\begin{equation}
D(\hat{r} /\hat{d})= \int_{-1/\hat{d}}^{1/\hat{d}} \frac{ d
\hat{Q}}{2}
\exp(i \hat{Q} \pi \hat{r}) \left[ I^d_0(\hat{Q} \pi \hat{d} )
-1  \right] .
\end{equation}
The theorem of Mehta and Mehta asserts the equality of the scaled
correlation functions for $\beta =2,4$ for all integer $r$, i.e. for
$\hat{r} = \hat{d} \times \mbox{integer}$
\begin{equation}
D(r \rightarrow \hat{r}/ \hat{d}) = C( r \rightarrow \hat{r}/ d).
\end{equation}

\subsection{Other Moments}

As opposed to the case when particles are in the continuum, the moment
$I_{1}(Q)$ is not interaction independent for a system where the
particles sit on a lattice: this is a well-known effect of the lattice
systems with interaction \cite{Sh-Su-90}. However, we can work
out an expression for the $1/r^{2}$ spin chain, using the usual
definition as a double commutator. In the remaining part we will
assume that $\hat{d} = 1/2$ and write
\begin{equation}
I_{1}(Q)=  \langle [[\hat{S_Q^z}, H], \hat{S}_{-Q}^z] \rangle.
\end{equation}
Calculating the commutator, we find
\begin{eqnarray}
I_{1}^{d}(Q) & = & \frac{2}{L} \sum_{i,j} (- J_{i,j}) [1-
\cos(Q(r_i-r_j))] \langle \hat{S}_i^z \hat{S}_j^z\rangle \nonumber \\
& = & - 2 J \phi^2 \sum_{r \neq 0} \frac{1}{\sin^2(\phi r)} [1-\cos(Q
r)]\langle \hat{S}_{0}^{z} \hat{S}_{r}^{z}\rangle.
\end{eqnarray}
Using Eq.~(\ref{eq:I0d}), we rewrite this as
\begin{equation}
I_{1}{^d}(Q) = \frac{\pi J}{ 2 L} \sum_{|k| \leq \pi} \ln(1- |k|/ \pi)
\sum_{r\neq 0} \left[ \frac{\phi^2}{\sin^2(\phi r)}\right] [1-\cos(k
r)]\cos(k r).
\end{equation}
Using the fact that, for $|k| \leq \pi$ \cite{Haldane},
\begin{equation}
\sum_{r \neq 0} \left[\frac{\phi^2}{\sin^2(\phi r)}\right] \cos(k r)
= \frac{\pi^2}{3} (1-1/L^2) - \pi |k| \left(1 - \frac{ |k|}{ 2
\pi}\right) \ ,
\end{equation}
we find with $[k] \equiv (k - 2 \pi m) \ [m \ \mbox{integer} | |[k]|
\leq \pi]$
\begin{eqnarray}
I_{1}^{d}(Q) & = & \frac{\pi J}{ 4 L} \sum_{|k| \leq \pi} \ln\left(1-
\frac{|k|}{\pi}\right) \Biggl[ |[Q +k]| + |[Q-k]| -2 |k| \nonumber \\
& - & \frac{1}{2 \pi} \{ [Q+k]^2+[q-k]^2 -2 k^2 \} \Biggr].
\end{eqnarray}
After turning the sum to an integral, we can integrate the expression
and find, for $ |Q|<\pi $,
\begin{equation}
I_{1}^{d}(Q) = \frac{J\pi^{2}}{8} \left[ 1 - \left( 1
- \frac{|Q|}{\pi} \right)^{2} + 2 \left( 1 - \frac{|Q|}{\pi}
\right)^{2} \ln \left( 1 - \frac{|Q|}{\pi} \right)\right] \ .
\label{eq:I1d}
\end{equation}
In the limit $Q \rightarrow 0$, we see that $I_{1}^{d}(Q) \rightarrow
J Q^2/4$ but for larger $Q$ there is substantial departure from the
pure quadratic behavior of the continuum models
Eq.~(\ref{eq:f-sumrule}).

In the hydrodynamic limit the moment $I_{-1}(Q)$ can be obtained from
the spin susceptibility $\chi_{spin}$, which is known explicitly
\cite{Shastry92}:
\begin{equation}
I_{-1}^{d}(Q) \stackrel{Q\rightarrow 0}{\longrightarrow} \left[
L\left(\frac{\partial^{2}E_{0}}{\partial {\cal M}^{2}}\right)_{{\cal
M}=0}\right]^{-1} = \frac{\chi_{spin}(Q\rightarrow 0)}{L} =
\frac{1}{\pi^{2}J} \ ,
\label{eq:I_1d}
\end{equation}
where ${\cal M}=(\hat{N}_{\uparrow}-\hat{N}_{\downarrow})/L$ is the
magnetization and the operators $\hat{N}_{\uparrow\downarrow}$ count
the number of up or down spins in the chain (notice that we have
$\hat{N}_{\uparrow}+\hat{N}_{\downarrow}=L/2)$.

With these moments at hand, we can define characteristic frequencies
for the excitation spectrum, in the same way as we did before. The
important point is that all those frequencies will have the same value
as $Q\rightarrow 0$:
\begin{equation}
\frac{I_{1}^{d}(Q)}{I_{0}^{d}(Q)} \stackrel{Q\rightarrow
0}{\longrightarrow} \frac{I_{0}^{d}(Q)}{I_{-1}^{d}(Q)}
\stackrel{Q\rightarrow 0}{\longrightarrow}
\sqrt{\frac{I_{1}^{d}(Q)}{I_{-1}^{d}(Q)}} \stackrel{Q\rightarrow
0}{\longrightarrow} J\pi\frac{|Q|}{2} \ .
\label{eq:limit}
\end{equation}
This means that in the hydrodynamic limit the spectrum is exhausted by
the excitations with dispersion relation $\omega=J\pi |Q|/2$
\cite{Haldane,Shastry88}. Although this might be taken as a common
feature for continuum systems, this is not so in general, as we shall
see below.

We see that, by choosing appropriate energy scale ($J \rightarrow
\frac{16}{\pi^2}$), the $1/r^{2}$ discrete spin model moments map
exactly onto the continuum symplectic ones for $Q\rightarrow 0$. The
dimensionless moments are identical in this limit,
\begin{equation}
I_n^d(Q=\pi\bar{q}/2) \stackrel{Q\rightarrow 0}{\longrightarrow}
\frac{1}{ k^n_F} \; I_n^s(q= k_F \bar{q}),
\end{equation}
where the factor of $1/2$ arises from the different normalization in
Eqs.~(\ref{eq:sqw},\ref{eq:sqw_disc}). It would be very interesting to
pursue the calculation of more moments at $q>0$ for both models to
check whether the discrete one shares the same unusual characteristics
of the continuum ($\beta=4$) model excitation spectrum.

We now turn to the Bethe chain, which is defined by the Hamiltonian $H
= J_H \sum_{i} \vec{S}_{i}\cdot\vec{S}_{i+1}$. This model was studied
in Ref.~\cite{Hohenberg}, and does not show the saturation at
$Q\rightarrow 0$. The moments $I_{1}$ and $I_{-1} $ are known
\cite{Hohenberg}, but not $I_{0}$:
\begin{equation}
I_{1}^{B}(Q) = \frac{J}{4} (2\ln2 -1/2)(1-\cos Q) \ ,
\label{eq:I1B}
\end{equation}
and
\begin{equation}
I_{-1}^{B}(Q) \stackrel{Q\rightarrow
0}{\longrightarrow} \frac{1}{\pi^{2} J} \ .
\label{eq:I_1B}
\end{equation}
The spinon spectrum of this system is known from Faddeev and
Takhtajan's work \cite{Faddeev}, $\omega_{sp}(Q)=(\pi J/2)\sin|Q|$.
Therefore, we could look for saturation by the spinon, and hence form
the ratio
\begin{equation}
\lim_{Q\rightarrow 0} \left[ \frac{1}{\omega_{sp}(Q)}
\sqrt{\frac{I_{1}^{B}(Q)}{I^{B}_{-1}(Q)}} \ \right] \approx 1.08706 \ .
\label{eq:insat}
\end{equation}
It is therefore clear that the small $Q$ behavior of $S(Q,\omega)$
{\it is not} exhausted by the spinons in the Bethe chain, unlike in
the $1/r^2$ model, where {\it it is}.

\section{Conclusions}
\label{sec:conclusions}

We have seen that the results of the calculation of Simons {\it et al}
\cite{Ben_Lee_Boris_PRL,Ben_Lee_Boris_PRB,Ben_Boris} for the
dynamical structure function has a representation in terms of the
Bethe quasi-particle quasi-hole energies. The representation obtained
in this work, has the character of two particle-hole pairs
representing the bare density fluctuation. We should note, however,
that this representation is far from unique, for one thing one may add
an arbitrary number of ``zero pairs''. Also, for example, we could
decompose our two holes and a particle as three holes and two
particles, by e.g. forcing the momenta of two holes to coincide, and
by suitably restricting the momenta. However, it is not possible to
decompose the results into those of a single particle-hole pair; our
representation in terms of two pairs appears to be minimal in some
sense. The striking feature which underlines the dynamical structure
factors presented in this paper is the truncation at very low orders of
series like Eq.~(\ref{eq:landau}). Therefore, the excited states for
system with $\beta=1,2$, and $4$ will always involve a small number of
quasi particle-hole pairs. This may not be true for arbitrary values
of the coupling constant.

We have also shown that the discrete model shares the property of
saturation of the Feynman sum rule by the lowest mode as $q
\rightarrow 0$. The static structure function obtained by a direct
calculation \cite{Vollhardt} is shown to be consistent with the
older calculation of Mehta's Ref.~\cite{2Mehtas} provided one
interprets the weight outside the Brillouin Zone appropriately by
Umklapping it. The first moment of the discrete model is obtained
using the known result for the two-point static correlator, and shows
interesting structure and departure from the first moment of the
continuum model, and should provide a non trivial check on the
structure function of the discrete model.

We stress that the saturation of the structure function by the sound
modes at small $q$ is a general property characterizing this class of
models.

\begin{center}
{\large Acknowledgment}
\end{center}

We are grateful to P.~A.~Lee, B.~Sutherland, and N.~Taniguchi for
useful discussions. This work was supported through NSF Grant No.~DMR
92-04480. B.D.S. wishes to acknowledge the financial support of the
SERC and NATO and E.R.M. wishes to thank CNPq for the financial
support.

\figure{1. Characteristic frequencies of $S(q,\omega)$. The solid line
corresponds to the Feynman spectrum $\omega_{{\cal F}}(q)$ and the
dashed line corresponds to the Hydrodynamical spectrum $\omega_{{\cal
H}}(q)$: (a) $\beta=2$; (b) $\beta=1$; (c) $\beta=4$.  \label{fig1}}

\figure{2. Region where $S^{u}(q,\omega)\neq 0$ ($\beta=2$). The
equations for the boundaries are indicated.  \label{fig2}}

\figure{3. Tridimensional plot of $S(q,\omega)$ ($\beta=2$) for the
unitary case. The vertical axis has a linear scale.  \label{fig3}}

\figure{4. Moments of $S(q,\omega)$ for $\beta=2$: (a) $I_{1}(q)$;
(b) $I_{0}(q)$; (c) $I_{-1}(q)$.  \label{fig4}}
\figure{5. Region where $S^{o}(q,\omega)\neq 0$ ($\beta=1$). The
equations for the boundaries are indicated.  \label{fig5}}

\figure{6. Tridimensional plot of $S(q,\omega)$ for the orthogonal
case ($\beta=1$). The delta function of Eq.~(\ref{eq:orthfunc})
has been regularized by a Gaussian. The vertical axis has a
logarithmic scale.  \label{fig6}}

\figure{7. Moments of $S(q,\omega)$ for $\beta=1$: (a) $I_{1}(q)$; (b)
$I_{0}(q)$; (c) $I_{-1}(q)$.  \label{fig7}}

\figure{8. Region where $S^{s}(q,\omega)\neq 0$ ($\beta=4$). The
equations for the boundaries are indicated.  \label{fig8}}

\figure{9. Tridimensional plot of $S(q,\omega)$ for the symplectic
case ($\beta=4$). The delta function of Eq.~(\ref{eq:sympfunc})
has been regularized by a Gaussian. The vertical axis has a
logarithmic scale.  \label{fig9}}

\figure{10. Moments of $S(q,\omega)$ for $\beta=4$: (a) $I_{1}(q)$;
(b) $I_{0}(q)$; (c) $I_{-1}(q)$.  \label{fig10}}

\figure{11. The two quasi-particle quasi-hole pair scheme for the
orthogonal and symplectic cases. `X' denotes particles and `O'
denotes holes, and the solid line indicates the range $-1 \leq k \leq
1$, i.e. the Fermi sea ($k_{F}=1$). The symplectic ensemble has two
pieces: case (a) corresponding to $S_a$ and case (b) corresponding
to $S_b$.  \label{fig11}}


\begin{thebibliography}{10}

\bibitem{Calogero} F. Calogero, J.~Math.~Phys. {\bf 10}, 2191,
2197 (1969).

\bibitem{Sutherland} B.~Sutherland, J.~Math.~Phys. {\bf 12}, 246,
251 (1971); Phys.~Rev.~A {\bf 4}, 2019 (1971); {\it ibid.} {\bf 5},
1372 (1972).

\bibitem{Dyson} F. J. Dyson, J. Math. Phys. {\bf 3}, 140, 157,166,
1191 (1963).

\bibitem{Haldane} F.~D.~M.~Haldane, Phys.~Rev.~Lett. {\bf 60},
635 (1988).

\bibitem{Shastry88} B.~S.~Shastry, Phys.~Rev.~Lett. {\bf 60},
639 (1988).

\bibitem{Sh-Su-90} B.~.S.~Shastry and B.~Sutherland,
Phys.~Rev.~Lett. {\bf 65}, 243 (1990).

\bibitem{Sutherland92} B.~Sutherland, Phys.~Rev.~B {\bf 45},
907 (1992).

\bibitem{Shastry92} B.~S.~Shastry, Phys.~Rev.~Lett.
{\bf 69}, 164 (1992).

\bibitem{Sh-Su-93} B.~S.~Shastry and B.~Sutherland,
Phys.~Rev.~Lett. {\bf 70}, 4029 (1993).

\bibitem{Su-Sh-93} B.~Sutherland and B.~S.~Shastry
Phys.~Rev.~Lett. {\bf 71}, 5 (1993).

\bibitem{Ben_Lee_Boris_PRL} B.~D.~Simons, P.~A.~Lee, and
B.~L.~Altshuler, Phys.~Rev.~Lett. {\bf 70}, 4122 (1993).

\bibitem{Ben_Lee_Boris_PRB} B.~D.~Simons, P.~A.~Lee, and
B.~L.~Altshuler (to appear in Phys.~Rev.~B).

\bibitem{Ben_Boris} B.~D.~Simons and B.~L.~Altshuler,
Phys.~Rev.~Lett. {\bf 70}, 4063 (1993); Phys.~Rev.~B
{\bf 48}, 5422 (1993).

\bibitem{Narayan} O.~Narayan and B.~S.~Shastry (to appear in
Phys.~Rev.~Lett.).

\bibitem{Ben_Lee_Boris_NPB} B.~D.~Simons, P.~A.~Lee, and
B.~L.~Altshuler (to appear in Nucl.~Phys.~B).

\bibitem{Sutherland_course} B. Sutherland, in {\it Exactly Solvable
Problems in Condensed Matter and Relativistic Field Theory}, edited by
B.~S.~Shastry {\it et al.}, Lecture Notes in Physics (Springer-Verlag,
Berlin, 1985), Vol. 242.

\bibitem{Pines} D.~Pines and P.~Nozi\`{e}res, {\it The Theory of
Quantum Liquids} (Addison-Wesley, Redwood City, 1989), Vol. I,
Chap. II.

\bibitem{Mehta} M.~L.~Mehta, {\it Random Matrices}, 2nd. ed.
(Academic Press, San Diego, 1991).

\bibitem{Feynman} R.~P.~Feynman, in {\it Progress in Low Temperature
Physics}, edited by C.~J.~Gorter (North-Holland, Amsterdam, 1957),
Vol. I, Chap. II.

\bibitem{mehtacorrs} M.~L.~Mehta in \cite{Mehta}, Eqs.~(5.2.20,
6.4.14 and 7.2.6) for $\beta=$ 2, 1, and 4, respectively.

\bibitem{Efetov} K.~B.~Efetov, Adv. in Phys. {\bf 32},
53 (1983).

\bibitem{Landau} L.~D.~Landau, Sov.~Phys.~JETP {\bf 30},
1058 (1956).

\bibitem{2Mehtas} M.~L.~Mehta and G.~C.~Mehta, J.~Math.~Phys.
{\bf 16}, 1256 (1975).

\bibitem{Vollhardt} F.~Gebhard and D.~Vollhardt, Phys.~Rev.~Lett.
{\bf 59}, 1472 (1987).

\bibitem{Hohenberg} P.~C.~Hohenberg and W.~F.~Brinkman, Phys.~Rev.~B
{\bf 10}, 128 (1974).

\bibitem{Faddeev} L.~Faddeev and L.~Takhtadjan, Phys.~Letts.~A
{\bf 85}, 375 (1981).

\end{thebibliography}
\end{document}